\documentclass[11pt]{article}
\usepackage{cospar}
\usepackage{natbib}
\usepackage{url}

\newcommand{\pub}[5] {#5, {\it #1}, {\bf #2}, #3, #4.}
\newcommand{\puba}[5]{#5, {\it #1}, #3, #4.}

\newcommand{\gray}{$\gamma$-ray}

\newcommand{\Dpp}{D_{pp}}
\newcommand{\Dxx}{D_{xx}}
\newcommand{\ddp}{\frac{\partial}{\partial p}}

\newcommand{\hi}{H {\sc i}}

\newcommand\paperno{
   \vspace{-17\baselineskip} \hspace{-0.05\textwidth}
   \begin{minipage}[t]{100mm}
   \noindent \bf H0.2-E0.2-0036-02 \rm(Proc.~COSPAR-2002, Houston, TX)\\
\bf LANL Report \# LA-UR-02-7812\\
\small \rm Submitted to \it Advances in Space Research
   \end{minipage}\vspace{15\baselineskip}}
\newcommand{\aap}{Astron.\ Astrophys.}
\newcommand{\adv}{Adv.\ Space Res.}
\newcommand{\apj}{Astrophys.\ J.}

\newcommand{\prc}{Phys.\ Rev.\ C}
\newcommand{\prl}{Phys.\ Rev.\ Lett.}

\title{PROPAGATION MODEL FOR COSMIC RAY SPECIES IN THE GALAXY}

\author{V.S. Ptuskin\address{IZMIRAN, Troitsk, Moscow region 142190, Russia},
        I. V. Moskalenko\address{NASA/Goddard Space Flight Center,
        Code 661, Greenbelt, MD 20771, U.S.A.}\address{JCA/University 
        of Maryland, Baltimore County, Baltimore, MD 21250, U.S.A.},
        F.C. Jones$^2$,
        A. W. Strong\address{MPI f\"ur extraterrestrische Physik, 
        Postfach 1603, D-85740 Garching, Germany},
        and
        S. G. Mashnik\address{Los Alamos National Laboratory, Los Alamos,
        NM 97545, U.S.A.}}

\begin{document}

\maketitle
\paperno

\begin{abstract}
In a recent paper \citep{M02}, it has been shown that 
the flux of secondary cosmic ray (CR) antiprotons appears to
be contradictory to measurements of secondary to primary
nuclei ratios in cosmic rays when calculated in the same 
Galactic propagation model.
The contradiction appears as
a value of the diffusion coefficient necessary to match
the secondary ratios $\bar p/p$ and B/C.
In particlular, it was shown that the reacceleration models designed to match
secondary to primary nuclei ratios produce too few
antiprotons. 
It is, however, clear that some reacceleration is
unavoidable in the turbulent interstellar medium.
Here we discuss an idea of how to improve reacceleration model
by allowing for the damping of interstellar turbulences on the small scale
by cosmic rays, mostly protons.
This would lead to increase in the mean free path lenghts at low energies,
the well-known phenomena empirically discovered in the Leaky Box models,
thus producing less secondary nuclei. 
Antiprotons will remain almost non-affected
due to their high energy threshold of production cross section.

\end{abstract}


\section*{INTRODUCTION}\label{sec:intro}

The spectrum and origin of antiprotons in CR has been a matter of active
debate since the first reported detections in balloon flights 
\citep{Golden79,bogomolov1}. There is a consensus that most of
the CR antiprotons observed near the Earth are ``secondaries'' produced in
collisions of energetic CR particles with interstellar gas
\citep[e.g.,][]{mitchell}.

The spectrum of secondary antiprotons has a peak at about 2 GeV
decreasing sharply towards lower energies. This unique shape
distinguishes antiprotons from other CR species.  
Over the last few years
the accuracy has been improved sufficiently
\citep[BESS 1995--2000,][]{Orito00,Sanu00,Asaoka02} that we can 
restrict the spectrum of the secondary component accurately enough to
test Galactic CR propagation models, and the heliospheric modulation.

It has been recently shown (\citealt{M01,M02}; see also \citealt{simon,sina})
that accurate antiproton
measurements during the last solar minimum 1995--1997
\citep[BESS,][]{Orito00} are inconsistent with existing propagation models
at the $\sim40$\% level at about 2 GeV while the stated measurement 
uncertainties in this energy range are now $\sim20$\%.
The conventional models based on local CR measurements,
simple energy dependence of the diffusion coefficient,  and uniform CR
source spectra throughout the Galaxy fail to reproduce
simultaneously both the secondary to primary nuclei ratio
and antiproton flux.

The reacceleration model designed to match secondary to primary nuclei
ratios, e.g., boron/carbon \citep{simon86,seo}
produce too few antiprotons because, e.g.,
matching the B/C ratio at all energies requires the diffusion
coefficient to be too large.  The models without reacceleration can
reproduce the antiproton flux, however they fall short of explaining the
low-energy decrease in the secondary to primary nuclei ratio.  To be
consistent with both, the introduction of breaks in the
diffusion coefficient and the injection spectrum is required, which
would suggest new phenomena in particle acceleration and propagation.

This forces us to developing a more sophisticated treatment of wave-particle
interaction in a course of CR propagation in the turbulent
interstellar medium. (An alternative idea of a local ``unprocessed''
nuclei component in low-energy CR is evaluated in \citealt{M02a}.)
In particular, we suggest a new approach which includes
the interaction of particles
with waves in the interstellar medium in a self-consistent way
and take into account the wave damping on energetic particles.
This requires numerical methods and iterative procedure
to derive the diffusion coefficient at arbitrary energy.
Secondary-to-primary elements ratios and antiproton flux should be used as final indicators
of the consistency.  
In our calculations we use CR propagation code GALPROP. 
This work is currently in progress.

\section*{BASIC FEATURES OF THE GALPROP MODELS} \label{sec:descr}

The cylindrically symmetric GALPROP models have been described in
detail elsewhere \citep{SM98}; here we summarize their basic features.

The models are three dimensional with cylindrical symmetry in the
Galaxy, and the basic coordinates are $(R,z,p)$ where $R$ is
Galactocentric radius, $z$ is the distance from the Galactic plane and
$p$ is the total particle momentum. In the models the propagation
region is bounded by $R=R_h$, $z=\pm z_h$ beyond which free escape is
assumed.

The propagation equation we use for all CR species is written in the form:
\begin{equation}
\label{eq.1}
\frac{\partial \psi}{\partial t} 
= q({\mathbf r}, p)+  \nabla  \cdot ( \Dxx\nabla\psi - {\mathbf V}\psi )
+ \ddp\, p^2 \Dpp \ddp\, \frac{1}{p^2}\, \psi
 -  \frac{\partial}{\partial p} \left[\dot{p} \psi
- \frac{p}{3} \, (\nabla \cdot {\mathbf V} )\psi\right]
- \frac{1}{\tau_f}\psi - \frac{1}{\tau_r}\psi\, ,
\end{equation}
where $\psi=\psi ({\mathbf r},p,t)$ is the density per unit of total
particle momentum, $\psi(p)dp = 4\pi p^2 f({\mathbf p})$ in terms of
phase-space density $f({\mathbf p})$, $q({\mathbf r}, p)$ is the source term,
$\Dxx$ is the spatial diffusion coefficient, ${\mathbf V}$ is the
convection velocity, reacceleration is described as diffusion in
momentum space and is determined by the coefficient $\Dpp$,
$\dot{p}\equiv dp/dt$ is the momentum loss rate, $\tau_f$ is the time
scale for fragmentation, and $\tau_r$ is the time scale for
radioactive decay. The numerical solution of the transport equation is
based on a Crank-Nicholson \citep{Press92} implicit second-order
scheme. The three spatial boundary conditions $\psi(R_h,z,p) =
\psi(R,\pm z_h,p) = 0$ are imposed on each iteration, where we take
$R_h=30$ kpc.

For a given $z_h$ the diffusion coefficient as a function of momentum
and the reacceleration or convection parameters is determined by
boron-to-carbon (B/C) ratio data. The spatial diffusion coefficient
is taken as $\Dxx = \beta D_0(\rho/\rho_0)^{\delta}$ if necessary with
a break ($\delta=\delta_1$ below rigidity $\rho_0$, $\delta=\delta_2$
above rigidity $\rho_0$). The injection spectrum of nucleons is
assumed to be a power law in momentum, $dq(p)/dp \propto p^{-\gamma}$
for the injected particle density.
For the case of reacceleration the momentum-space
diffusion coefficient $D_{pp}$ is related to the spatial coefficient
$\Dxx$ \citep{berezinskii,seo} via the Alfv\'en speed $v_A$.
The convection velocity (in $z$-direction only) $V(z)$ is assumed to
increase linearly with distance from the plane ($dV/dz>0$ for all
$z$). This implies a constant adiabatic energy loss. 

The interstellar hydrogen distribution uses \hi\ and CO surveys
and information on the ionized component; the helium fraction of the
gas is taken as 0.11 by number. The H$_2$ and \hi\ gas number
densities in the Galactic plane are defined in the form of tables,
which are interpolated linearly. The extension of the gas distribution
to an arbitrary height above the plane is made using analytical
approximations.

The distribution of CR sources is chosen to reproduce the
CR distribution determined by analysis of EGRET \gray\ data
\citep{StrongMattox96} and was described in \citet{SM98}.

Energy losses for nucleons by ionization and Coulomb interactions are
included, and for electrons by ionization, Coulomb interactions,
bremsstrahlung, inverse Compton, and synchrotron.

Positrons and electrons (including secondary electrons) are propagated
in the same model. Positron production is computed as described in
\citet{MS98}, that paper includes a critical reevaluation of the
secondary $\pi^\pm$- and $K^\pm$-meson decay calculations.

Gas-related \gray\ intensities are computed from the emissivities as a
function of $(R,z,E_\gamma)$ using the column densities of \hi\ and
H$_2$. The interstellar radiation field, used for calculation
of the inverse Compton emission and electron energy losses, is
calculated based on stellar population models and COBE results, plus
the cosmic microwave background.

\subsection*{New Developments} \label{sec:new}

The experience gained from the original fortran--90 code allowed us to
design a new version of the model, entirely rewritten in C++, that is
much more flexible. It allows essential optimizations in comparison
to the older model and a full 3-dimensional spatial grid. It is now
possible to explicitly solve the full nuclear reaction network on a
spatially resolved grid. The code can thus serve as a complete
substitute for the conventional ``leaky-box'' or ``weighted-slab''
propagation models usually employed, giving many advantages such as
the correct treatment of radioactive nuclei, realistic gas and source
distributions etc. It also allows stochastic SNR sources to be
included. It still contains an option to switch to the fast running
cylindrically symmetrical model which is sufficient for many
applications such as the present one.

In the new version, we have updated the cross-section code to include
the latest measurements and energy dependent fitting functions. The
nuclear reaction network is built using the Nuclear Data Sheets. The
isotopic cross section database consists of thousands of points
collected from the literature. This includes a
critical re-evaluation of some data and cross checks. The isotopic
cross sections are calculated using the author's fits to major
beryllium and boron production cross sections $p+\mathrm{C,N,O} \to \mathrm{
Be,B}$. Other cross sections are calculated using
phenomenological approximations by \citet{W-code} and/or 
\citet{ST-code} renormalized to the data
where it exists. The cross sections on the He target are calculated
using a parametrization by \citet{ferrando88}.

The reaction network is solved starting at the heaviest nuclei (i.e.,
$^{64}$Ni). The propagation equation Eq.~(\ref{eq.1}) is solved, computing all the
resulting secondary source functions, and then proceeds to the nuclei
with $A-1$. The procedure is repeated down to $A=1$. In this way all
secondary, tertiary etc.\ reactions are automatically accounted for.
To be completely accurate
for all isotopes, e.g., for some rare cases of $\beta^\pm$-decay, the
whole loop is repeated twice. Our preliminary results for all CR
species $Z\leq28$ are given in \citet{SM01}.

\section*{COSMIC RAYS IN INTERSTELLAR TURBULENCE: A SELF-CONSISTENT APPROACH}

It is known that galactic CR have relatively high energy density and
they can not always be treated as test particles moving in given interstellar
magnetic fields, see \citet{berezinskii}. In particular, when the
stochastic reacceleration of CR is considered, one has to take into
account the damping of waves, which loose energy on particle acceleration. The
damping causes the change of the wave spectrum that in its turn affects the
particle transport. Thus in principle the study of CR propagation may
need a self-consistent consideration. We shall see from the numerical estimates
below that the back reaction of CR on interstellar turbulence should
be taken into account at particle energies below 10 GeV/nucleon.

It is assumed that the wave-particle interaction in the interstellar plasma
has a resonant character (the cyclotron resonance). The scattering of
energetic particles by random hydromagnetic waves leads to the spatial
diffusion of CR in the Galaxy. The CR diffusion coefficient
is determined by the following approximate equation \citep{berezinskii}:
\begin{equation}
\label{eq.2}
D(p)=\frac{vr_{g}B^{2}}{12\pi k_{res}W(k_{res})},
\end{equation}
where $v$ is the particle velocity, $r_{g}=pc/\left(ZeB\right)  $ is the
particle Larmor radius in the average magnetic field $B$ ($Ze$ is the particle
charge), $k_{res}=1/r_{g}$ is the resonant wave number, and $W(k)$ is the
spectral energy density of waves defined as $\int dkW(k)=\delta B^{2}/4\pi$
($\delta B$ is the random magnetic field, $\delta B\ll B$).

In its simplified form, the steady state equation for hydromagnetic waves with
a nonlinear energy transfer in $k$-space can be written as
\begin{equation}
\label{eq.3}
\frac{\partial}{\partial k}\left(  C_{A}k^{2}W(k)\sqrt{\frac{kW(k)}{4\pi\rho}%
}\right)  =-2\Gamma_{cr}(k)W(k)+S\delta(k-k_{L}),
\end{equation}
$k\geq k_{L}$ \cite[e.g.,][]{landau,norman}. Here the l.h.s.\ of equation
describes the Kolmogorov type nonlinear cascade from small $k$ to large $k$,
$C_{A}$ is a constant and according to the numerical simulations of
magnetohydrodynamical turbulence by \citet{verma}
the magnitude of $C_{A}$ is very roughly equal to $0.3$, $\rho$ is the
interstellar gas density. The r.h.s.\ of Eq.~(\ref{eq.3}) includes the wave damping
on CR and the source term which works on a large scale $\sim1/k_{L}$
and describes the generation of turbulence by supernova bursts, powerful
stellar winds, and superbubbles expansion. In the limit of negligible damping,
$\Gamma_{cr}=0$, the solution of Eq.~(\ref{eq.3}) gives the Kolmogorov spectrum
$W(k)\propto k^{-5/3}$ at $k>k_{L}$. The latter leads to the diffusion
coefficient $D(p)\propto v(p/Z)^{1/3}$.

The expression for wave amplitude attenuation is given by \citep{berezinskii}:
\begin{equation}
\label{eq.4}
\Gamma_{cr}=\frac{\pi e^{2}V_{a}^{2}}{2kc^{2}}\int_{p_{res}(k)}^{\infty}%
\frac{dp}{p}\Psi(p),
\end{equation}
where $V_{a}=B/\sqrt{4\pi\rho}$ is the Alfven velocity, $p_{res}=ZeB/ck$ is
the resonant momentum.

The solution of Eqs.~(\ref{eq.3}), (\ref{eq.4}) at $k>k_{L}$ gives
\begin{equation}
\label{eq.5}
\frac{kW(k)}{B^{2}}=\left[  \left(  \frac{k_{L}}{k}\right)  ^{1/3}\left(
\frac{k_{L}W(k_{L})}{B^{2}}\right)  ^{1/2}-\frac{\pi e^{2}V_{a}}{3C_{A}%
c^{2}k^{1/3}}\int_{k_{0}}^{k}dk_{1}k_{1}^{-8/3}\int_{p_{res}(k_{1})}^{\infty
}\frac{dp}{p}\Psi(p)\right]^{2}.
\end{equation}
The wave damping on CR is essential only at relatively small scales
$k^{-1}<k_{d}^{-1}$, $k_{d}\ll k_{L}^{-1}$ (in fact, $k_{d}^{-1}\sim10^{12}$
cm, $k_{L}^{-1}\sim10^{20}$ cm). The turbulence at $k_{d}^{-1}\ll k^{-1}%
<k_{L}^{-1}$ obeys the Kolmogorov scaling $W(k)\propto k^{-5/3}$. It is
clear from Eq.~(\ref{eq.5}) that the dissipation of waves on CR described
by the second term in square brackets makes the wave spectrum at large wave
numbers steeper than the Kolmogorov spectrum.

The diffusion mean free path $l(p)$, as defined by $D(p)=vl(p)/3$, is now
given by:
\begin{equation}
\label{eq.6}
l(p)=r_{g}^{2/3}(p_{L})r_{g}^{1/3}(p)\left[  \left(  \frac{r_{g}(p_{L}%
)}{l(p_{L})}\right)  ^{1/2}-\frac{2\pi^{3/2}V_{a}p_{L}^{2}}{3C_{A}B^{2}}%
\int_{p/p_{L}}^{1}dtt^{2/3}\int_{tp_{L}}^{\infty}\frac{dp_{1}}{p_{1}}%
\Psi(p_{1})\right]  ^{-2},
\end{equation}
where we use the resonant conditions $p=p_{res}(k)$ and introduce
$p_{L}=p_{res}(k_{L})$. Eq.~(\ref{eq.6}) is formally valid at $p<p_{L}\sim10^{17}$
eV/c. The second term in square brackets reflects the modification of diffusion
mean free path due to the wave dissipation on CR. It increases with
decreasing the particle energy. It is easy to estimate that the second term in
square brackets is small at high energies but it is comparable with the first
term in square brackets at $p\sim1$ GeV/c (at $l=3$ pc, $V_{a}=30$ km s$^{-1}$, $B=3$
$\mu$G). So, the nonlinear modification of CR diffusion is expected at
$E<10$ GeV/nucleon.

As the most abundant specie, the CR protons mainly determine the wave
dissipation. Their distribution function $\Psi(p)$ should be used to calculate
$W(k)$ and $l(p)$. The self-consistent treatment of CR propagation
implies the solution of transport Eq.~(\ref{eq.1}) for $\Psi(p)$ with the
diffusion mean free path Eq.~(\ref{eq.6}), which is a function of $\Psi(p)$.

To demonstrate the effect of wave damping, let us consider a simple case of
one-dimensional diffusion model of CR propagation with the source
distribution $q(r,p)=q_{0}(p)\delta(z)$ (that corresponds to the infinitely
thin disk of CR sources located at $z=0$) and the flat CR halo
of height $H$, see \citet{berezinskii}, \citet{jones01}. The CR
source spectrum is of the form $q_{0}(p)\propto p^{-\gamma_{s}}$ and
the index is estimated as $\gamma_{s}=2.0...2.4$.

Let us assume that stochastic reacceleration does not essentiallly change the
CR spectrum on the characteristic time of CR leakage from the
Galaxy. We also ignore ionization energy losses and losses for nuclear
fragmantation and assume that the equilibrium density of CR protons is
determined by the balance between their production in the sources and the free
diffusion leakage from the Galaxy. The solution of the diffusion equation for
the relativistic protons in the galactic disk is then
\begin{equation}
\label{eq.7}
\Psi(p)=\frac{3q_{0}(p)H}{2vl(p)}.
\end{equation}

We simplify Eq.~(\ref{eq.6}) by using the approximation 
$\int_{tp_{L}}^{\infty}
\frac{dp_{1}}{p_{1}}\Psi(p_{1})\approx\Psi(tp_{L})$ and write it down as
\begin{equation}
\label{eq.8}
l(p)\approx l_{K}(p)\left[  1-\frac{2\pi^{3/2}V_{a}p^{1/3}}{3C_{A}B^{2}%
}\left(  \frac{l_{K}(p)}{r_{g}(p)}\right)  ^{1/2}\int_{p}^{p_{L}}dp_{1}%
p_{1}^{2/3}\Psi(p_{1})\right]  ^{-2},
\end{equation}
where $l_{K}(p)$ is the diffusion mean free path in the turbulence with
Kolmogorov spectrum, which is not modified by the wave damping on CR.

Now one can obtain the following solution of Eqs.~(\ref{eq.7}) and (\ref{eq.8}) for the
diffusion mean free path in the case of a power law source spectrum:
\begin{equation}
\label{eq.9}
l(p)=l_{K}(p)\left(  1+\frac{\pi^{3/2}V_{a}Hp^{1/3}}{C_{A}B^{2}}\left(
\frac{l_{K}(p)}{r_{g}(p)}\right)  ^{1/2}\int_{p}^{p_{L}}dp_{1}p_{1}^{2/3}%
\frac{q_{0}(p_{1})}{v(p_{1})l_{K}(p_{1})}\right)  ^{2}%
\end{equation}
The last expression together with Eq.~(\ref{eq.7}) presents the self-consistent
solution of the considered simple problem.

The low-energy asymptotic mean free path Eq.~(\ref{eq.9}) is $l(p)\propto
v^{-2}p^{3-2\gamma_{s}}$. The high-energy asymptotic mean free path Eq.~(\ref{eq.9})
is $l(p)\approx l_{K}(p)\propto p^{1/3}$ that correspons to a Kolmogorov
spectrum of waves. One can check that the wave spectrum preserves the
Kolmogorov scaling $W(k)\propto k^{-5/3}$ at small $k$ and it is modified
as $W(k)\propto$ $k^{1-2\gamma_{s}}$ at large wave numbers.

\section*{CONCLUSION}

The back reaction of energetic particles on interstellar turbulence may change
the spectrum of hydromagnetic waves at large wave numbers that leads to the
change of CR diffusion coefficient. In particular, the modification of
Kolmogorov type spectrum of turbulence leads to the CR diffusion
coefficient which goes through a minimum at particle magnetic rigidity
about a few GV, so that the diffusion coefficient has the power law
asymptotics $D(p)\propto vp^{1/3}$ and $D(p)\propto$ $v^{-1}
p^{3-2\gamma_{s}}$ at large and small $p$ respectively (the exponent of CR
source spectrum $\gamma_{s}\sim2.2$). Such a behaviour reduces the rate of
stochastic reacceleration of CR in the interstellar medium that is
needed to reproduce the observerd peaks in the secondary-to-primary elements
ratios in CR at these rigidities. The reduced rate of reacceleration
may help to explain the antiproton data.

In a future work, we plan to undertake a full scale self-consistent modeling of
CR transport in the turbulent interstellar medium in the frameworks of
the realistic galactic model described here and with the damping of
waves on energetic particles included.

\section*{ ACKNOWLEDGEMENTS}
V.\ S.\ P.,
I.\ V.\ M., and S.\ G.\ M.\ acknowledge partial support from
NASA Astrophysics Theory Program grants. The work of V.\ S.\ P.\ was also supported by
RFBR-01-02-17460 grant at IZMIRAN.

\enlargethispage{\baselineskip}

\bf \bigskip
\smallskip\noindent
E-mail address of I.V.~Moskalenko: \hspace{2mm} {imos@milkyway.gsfc.nasa.gov}

\medskip\noindent
Manuscript received \hspace{30mm}; revised \hspace{30mm}; accepted

\end{document}